\documentclass[11pt,showpacs,preprintnumbers,superscriptaddress,amsmath,amssymb,nofootinbib]{revtex4}
\usepackage{graphicx}% Include figure files
\usepackage{dcolumn}% Align table columns on decimal point
\usepackage{bm}% bold math
\usepackage{amssymb}
\usepackage{amsmath}
\usepackage{epsfig}    
\usepackage{color}
\usepackage{hhline}
\def\be{\begin{equation}}
\def\ee{\end{equation}}
\newcommand{\bea}{\begin{eqnarray}}
\newcommand{\eea}{\end{eqnarray}}
\newcommand{\nn}{\nonumber}

\numberwithin{equation}{section}

\begin{document}

%%%%%%%%%
\title{Higgs Triplet Model with Classically Conformal Invariance}
\author{Hiroshi Okada}
\email{hokada@kias.re.kr}
\affiliation{School of Physics, KIAS, Seoul 130-722, Korea}
\affiliation{Physics Division, National Center for Theoretical Sciences, Hsinchu, Taiwan 300}

\author{Yuta Orikasa}
\email{orikasa@kias.re.kr}
\affiliation{School of Physics, KIAS, Seoul 130-722, Korea}
\affiliation{Department of Physics and Astronomy, Seoul National University, Seoul 151-742, Korea}

\author{Kei Yagyu}
\email{k.yagyu@soton.ac.uk}
\affiliation{School of Physics and Astronomy, University of Southampton, Southampton, SO17 1BJ, United Kingdom}

\date{\today}

\begin{abstract}

We discuss an extension of the minimal Higgs triplet model with a classically conformal invariance and with a gauged $U(1)_{B-L}$ symmetry. 
In our scenario, tiny masses of neutrinos are generated by a hybrid contribution from the type-I and type-II seesaw mechanisms. 
The shape of the Higgs potential at low energies is 
determined by solving one-loop renormalization group equations for all the scalar quartic couplings 
with a set of initial values of parameters at the Planck scale. 
We find a successful set of the parameters 
in which the $U(1)_{B-L}$ symmetry is radiatively broken via the Coleman-Weinberg mechanism at the ${\cal O}$(10) TeV scale, 
and the electroweak symmetry breaking is also triggered by the $U(1)_{B-L}$ breaking. 
Under this configuration, 
we can predict various low energy observables such as the mass spectrum of extra Higgs bosons, and the mixing angles. 
Furthermore, using these predicted mass parameters, we obtain upper limits on Yukawa couplings among an isospin 
triplet Higgs field and lepton doublets from lepton flavor violation data. 

\end{abstract}
\maketitle
\newpage

\section{Introduction}

The existence of the Higgs boson has been established by the LHC experiments~\cite{LHC_ATLAS,LHC_CMS}, which
has given us the important guideline to consider physics beyond the Standard Model (SM). 
Namely, the Higgs sector has to contain at least one physical neutral scalar particle with its mass of about 125 GeV whose
property is consistent with that of the Higgs boson in the SM~\cite{LHC_ATLAS2,LHC_CMS2}. 
This situation, however, does not necessarily exclude possibilities to consider non-minimal Higgs sectors, e.g., a model with additional isospin multiplet scalar fields. 
Such a non-minimal Higgs sector often appears in various new physics models beyond the SM, and its property depends on a new physics scenario. 
Therefore, reconstruction of the structure of the Higgs sector is quite important to narrow down  new physics models. 

On the other hand, the discovery of the Higgs boson gave us a good opportunity to 
seriously consider what is the origin of the negative mass term in the Higgs potential 
as it has been discussed for a long time. 
One of the excellent explanations for this issue was proposed in the famous paper by S.~Coleman and E.~Weinberg~\cite{CW}, in which 
all the dimensionful parameters are forbidden by a classical conformal invariance (CCI), and the negative mass term is generated by a quantum effect. 
However, it has also been well known that the Coleman-Weinberg mechanism does not work to realize the spontaneous 
electroweak symmetry breaking within the SM particle content, because of the too strong negative contribution of the top quark loop~\cite{Sher}. 
In order to have the successful electroweak symmetry breaking, we need to have additional positive bosonic loop contributions. 
Therefore, extensions of the bosonic sector, e.g., introducing additional scalar multiplets 
and/or extra gauge symmetries, are good match in the scenario with the CCI~\cite{conformal}. 

In this paper, we consider an extension of the minimal Higgs Triplet Model (HTM) with the CCI and with a gauged $U(1)_{B-L}$ symmetry. 
The HTM is one of the well motivated non-minimal Higgs sectors, because it gives a simple explanation of tiny neutrino masses~\cite{typeII}. 
In our model, the electroweak symmetry breaking is triggered by the radiative breaking of the $U(1)_{B-L}$ symmetry at an ${\cal O}(10)$ TeV scale 
via the Coleman-Weinberg mechanism. 
Majorana masses of left-handed neutrinos are then generated through a hybrid contribution~\cite{hybrid,Schmidt:2007nq} of the type-I~\cite{typeI} and type-II~\cite{typeII} seesaw mechanisms. 
In our scenario, low energy observables such as masses of Higgs bosons and mixings
can be predicted by using the one-loop renormalization group equations (RGEs) with a set of fixed initial values of model parameters at a high energy. 

This paper is organized as follows. 
In Sec.~II-A, we first explain the setup of our model, and give a particle content. 
We then investigate how the $B-L$ and the electroweak symmetries are successfully broken in Sec.~II-B. 
In Sec.~II-C, we discuss the lepton sector of our model especially focusing on the neutrino mass generation and lepton flavor violation (LFV) processes. 
In Sec.~II-D, we give the kinetic term Lagrangian for scalar fields. 
In Sec.~III, we numerically solve the one-loop RGEs, and give predictions of low energy observables.  
Conclusions are given in Sec.~IV. 
In Appendix, we present the analytic expressions for the one-loop beta functions of all the dimensionless couplings in our model.

%%%%%%%%%%%%%%%%%%%%%%%%%%%%%%%%%%%%%%
\section{The Model}\label{sec:model}

\subsection{Setup}

We consider an extension of the minimal HTM with a CCI. 
In the minimal HTM, a scalar trilinear interaction $\mu\Phi^T(i\tau_2)\Delta^\dagger \Phi $, where $\Phi(\Delta)$ is
an isospin doublet (triplet) field with the hypercharge $Y=1/2~(1)$
plays an important role to give Majorana masses for neutrinos at the tree level~\cite{typeII}. 
In our scenario, however, this term is forbidden due to the CCI, but it is effectively induced from a dimensionless coupling constant by introducing 
an additional isospin singlet scalar field $\varphi$ as
\begin{align}
\lambda_0 \varphi \Phi^T(i\tau_2)\Delta^\dagger \Phi. \label{eq:lam0}
\end{align}
Thus, after $\varphi$ gets a non-zero VEV, the $\mu$ term is effectively generated. 

The value of $\lambda_0$ coupling at an arbitrary scale $M$ is determined by using one-loop RGEs with 
a fixed initial value at an initial scale $M_0$. 
Naturally, the $\lambda_0$ coupling  
is given to be zero at a high energy, e.g., the Planck scale, because  
the quartic vertex in Eq.~(\ref{eq:lam0}) is expected to be forbidden by global symmetries. 
For example, at high energies, the following global symmetry is expected to be restored:
\begin{align}
\Phi   \to e^{i\tau^a\theta_\Phi^a}\Phi,\quad 
\Delta \to e^{-i\tau^a\theta_\Delta^a}\,\Delta\,e^{i\tau^a\theta_\Delta^a}, \label{global}
\end{align}
where $\tau^a$ ($a=1$-3) are the Pauli matrices, and $\theta_\Phi$ and $\theta_\Delta$ are the $SU(2)$ rotation angles. 
Therefore, if the model is invariant under the transformation of Eq.~(\ref{global}), the vertex in Eq.~(\ref{eq:lam0}) is forbidden\footnote{The 
ordinary isospin $SU(2)_L$ invariance is, of course, kept in Eq.~(\ref{eq:lam0}) which corresponds to the transformation in Eq.~(\ref{global}) with $\theta_\Phi = \theta_\Delta$.  }. 

\begin{figure}[t]
\begin{center}
\includegraphics[scale=0.6]{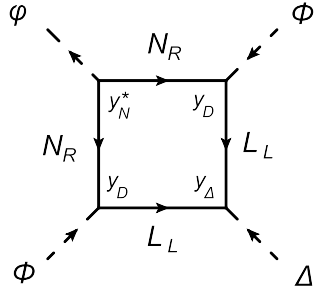}
\caption{The one-loop contribution to the $\lambda_0$ coupling, where the Yukawa coupling constants $y_N^{}$, $y_D^{}$ and $y_\Delta^{}$ 
are defined in Eq.~(\ref{Yukawa}). }
\label{fig:lam0}
\end{center}
\end{figure}

However, once we input $\lambda_0=0$ at high energies, the value of $\lambda_0$ is always zero at low energy scales, because the 
beta function for $\lambda_0$ is proportional to $\lambda_0$ itself at any loop levels as long as we consider the HTM with $\varphi$. 
In order to avoid such a situation, we introduce right-handed neutrinos, by which 
we obtain a term without proportional to $\lambda_0$ in the beta function from the diagram depicted in Fig.~\ref{fig:lam0}. 
It has been well known that right-handed neutrinos with three flavors can be naturally introduced in a model with a gauged $B-L$ symmetry~\cite{B-L} 
due to the gauge anomaly cancellation. 
We thus introduce the $U(1)_{B-L}$ gauge symmetry.  
In this case, when we assign a non-zero charge of $U(1)_{B-L}$ to $\varphi$, it can be identified as the Higgs field which is 
responsible to happen the spontaneous $U(1)_{B-L}$ symmetry breaking. 
Consequently, our CCI extended HTM is defined as shown in Table~\ref{tab:1} based on the $SU(2)_L\times U(1)_Y\times U(1)_{B-L}$ gauge theory.  

\begin{center}
\begin{table}[t]
\begin{tabular}{c||c|c|c||c|c|c}\hline\hline  
&\multicolumn{3}{c||}{Lepton Fields} & \multicolumn{3}{c}{Scalar Fields}  \\\hline
          &~$L_L$~ &~$e_R$~ &~$N_R$~ &~$\Phi$~  & ~$\Delta$& ~$\varphi$~   \\\hline 
$SU(2)_L$ & $\bm{2}$ & $\bm{1}$& $\bm{1}$ & $\bm{2}$&  $\bm{3}$ & $\bm{1}$ \\\hline 
$U(1)_Y$  & $-1/2$   & $-1$    & $0$     & $+1/2$    & $+1$   & $0$  \\\hline
$U(1)_{B-L} $ & $-1$ & $-1$ & $-1$ & $0$ & $2$ &  $2$    \\\hline\hline  
\end{tabular}
\caption{Contents of lepton and scalar fields
and their charge assignment under the gauge symmetry $SU(2)_L\times U(1)_Y\times U(1)_{B-L}$.
}
\label{tab:1}
\end{table}
\end{center}

\subsection{Higgs sector}

The most general form of the CCI Higgs potential under the $SU(2)_L\times U(1)_Y\times U(1)_{B-L}$ invariance is given by 
\begin{align}
V(\varphi,\Phi,\Delta)
&= 
 \lambda_1 (\Phi^\dag\Phi)^2 
  + \lambda_2 [{\rm Tr}(\Delta^\dag\Delta)]^2   + \lambda_3 {\rm Tr}[(\Delta^\dag\Delta)^2]
  + \lambda_4 (\Phi^\dag\Phi){\rm Tr}(\Delta^\dag\Delta)   +\lambda_5 \Phi^\dagger \Delta \Delta^\dagger \Phi  \nn\\
 &+\lambda_6 (\varphi^\dag\varphi)^2
  +\lambda_7  (\Phi^\dag\Phi)(\varphi^\dag\varphi)  
  +\lambda_8 {\rm Tr}(\Delta^\dag\Delta)(\varphi^\dag\varphi) 
+\left(\lambda_0 \varphi \Phi^Ti\tau_2 \Delta^\dag\Phi+{\rm h.c.}\right), 
\label{HP}
\end{align}
where $\lambda_0$ is taken real without loss of generality by rephasing the scalar fields. 
The scalar fields can be parameterized as
\begin{align}
\varphi = \frac1{\sqrt2}(\varphi_R^{}+v_\varphi +i\varphi_I^{}),\quad 
\Phi =\left(
\begin{array}{c}
\phi^+\\
\frac{\phi_R^{} +v_\phi+i\phi_I^{}}{\sqrt{2}}
\end{array}\right),\quad 
\Delta =\left(
\begin{array}{cc}
\frac{\Delta^+}{\sqrt2} & \Delta^{++}\\
 \frac{\Delta_R+v_\Delta+i \Delta_I}{\sqrt{2}}  & -\frac{\Delta^+}{\sqrt2}
\end{array}\right), 
\label{component}
\end{align}
where $v_\varphi$, $v_\phi$ and $v_\Delta^{}$ are the VEVs of the singlet, doublet and triplet scalar fields, respectively.  
At this stage, we do not discuss how the $B-L$ and the electroweak symmetry breaking occur, so that 
the non-zero VEVs for $\varphi$, $\Phi$ and $\Delta$ fields are not justified yet. 
In the following, we discuss the spontaneous breaking of the $B-L$ and electroweak symmetries. 

First, we investigate the spontaneous breakdown of the $U(1)_{B-L}$ symmetry.  
We assume that the VEV of $\varphi$ is given to be a muti-TeV scale, which is required by the constraint from the LEP experiments~\cite{LEP}. 
Because the magnitude of the VEVs of $\Phi$ and $\Delta$ is expected to be the electroweak scale, 
we can neglect $\left(\frac{v_\phi}{v_\varphi}\right)^2$ and $\left(\frac{v_\Delta}{v_\varphi}\right)^2$ terms. 
In this case, we can separately consider the $B-L$ sector and the other sector relevant to $\Phi$ and $\Delta$. 

The renormalization group improved effective potential for the $B-L$ sector is then given by 
\begin{eqnarray}
V_{B-L}(\tilde{\varphi})=\frac{\lambda_6(t)}{4} G^4(t) \tilde{\varphi}^4, 
\end{eqnarray}
where $\tilde{\varphi}=\sqrt{2}\text{Re}(\varphi)$ 
and $\lambda_6(t)$ with $t=\ln(\tilde{\varphi} /v_\varphi )$ is 
the scale dependent coupling which is evaluated by the one-loop beta 
function given in Eq.~(\ref{lam6}) in Appendix. 
The anomalous dimension $G(t)$ is given by
\begin{eqnarray}
G(t)= \exp \left[-\int_0^t dt' \gamma(t')\right], 
\end{eqnarray}
where the explicit form of $\gamma$ in the Landau gauge is 
\begin{eqnarray}
\gamma(t) =\frac{1}{32\pi^2}\left[ 4\text{Tr} (y_N^\dagger y_N ) -24g_{B-L}^2\right]. \label{sgamma}
\end{eqnarray}
In the above equation (\ref{sgamma}), $y_N^{}$ and $g_{B-L}^{}$ 
are the Yukawa coupling among the right-handed neutrinos and $\varphi$ defined in Eq.~(\ref{Yukawa}) and 
the $U(1)_{B-L}$ gauge coupling, respectively. 
The stationary condition at the scale $v_\varphi$ is given by 
\begin{eqnarray}
\left.\frac{dV_{B-L}}{d\tilde{\varphi}}\right|_{\tilde{\varphi}=v_\varphi}
=\frac{e^{-t}}{v_\varphi}
\left.\frac{dV_{B-L}}{dt}\right|_{t=0}
=0. 
\label{B-Ltad}
\end{eqnarray}
This equation leads to a relation among the renormalized coupling constants at the potential minimum such that 
\begin{eqnarray}
\frac{d\lambda_6}{dt} +4(1-\gamma)\lambda_6
\sim \frac{1}{16\pi^2}\left[20\lambda_6^2
+2\lambda_7^2+\lambda_8^2-8\text{Tr}(y_N^\dagger y_N y_N^\dagger y_N)+96 g_{B-L}^4\right]+4\lambda_6
=0. 
\end{eqnarray}
In the perturbative regime, i.e., 
$\lambda_6 \sim \lambda_{7,8}^2  \sim g_{B-L}^4 \sim y_N^4\ll 1$, 
we find a solution 
\begin{eqnarray}
\lambda_6(t=0) \sim   -\frac14\frac{1}{16\pi^2}\left[2\lambda_7^2+3\lambda_8^2
+96 g_{B-L}^4-8\text{Tr}(y_N^\dagger y_N y_N^\dagger y_N) \right]. \label{eq:tad1}
\end{eqnarray}
Thus, the $B-L$ breaking scale can be found by looking at the intersection point of 
the running of $\lambda_6$ and that of the right hand side of Eq.~(\ref{eq:tad1}).
The squared mass of $\varphi_R$ is then calculated at the $B-L$ breaking scale, i.e., $t=0$ as 
\begin{eqnarray}
m_\varphi^2=
\left.\frac{d^2V_{B-L}}{d\tilde{\varphi}^2}\right|_{\tilde{\varphi}=v_\varphi}
\sim -4\lambda_6(0) v_\varphi^2, \label{mphisq}
\end{eqnarray}
where we used $d\lambda_6/dt \simeq -4\lambda_6$ and $d^2\lambda_6/dt^2 \simeq g_{B-L}^2 d\lambda_6/dt \ll d\lambda_6/dt$. 
It is clearly seen that 
$\lambda_6(0)<0$ is required to have the correct sign of the mass term and to realize the 
spontaneous $B-L$ symmetry radiatively.
We can find a parameter space which satisfies $\lambda_6(0)<0$ as it will be shown numerically in Sec.~\ref{sec:RGE}.

\begin{figure}[t]
\begin{center}
\includegraphics[scale=0.8]{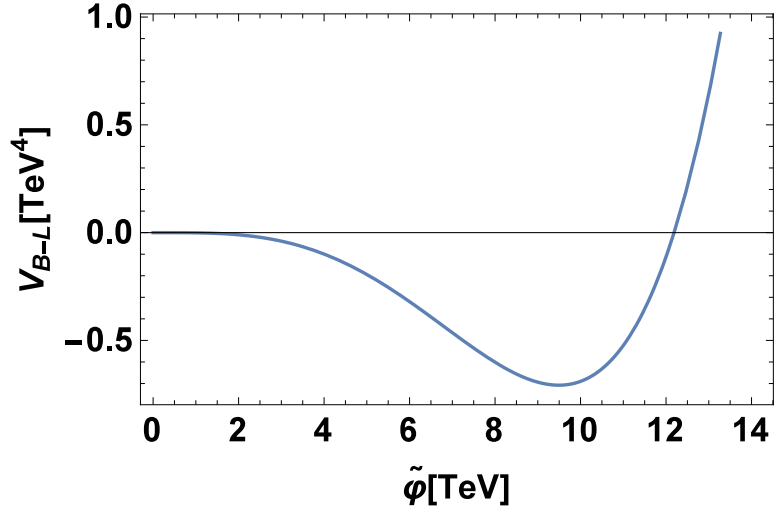} 
\caption{The renormalization group improved effective potential for the $B-L$ sector. 
We use the parameter set given in Table~\ref{initial}. }
\label{fig:potential}
\end{center}
\end{figure}

In Fig.~\ref{fig:potential}, we show the effective potential $V_{B-L}$ as a function of the order parameter $\tilde{\varphi}^{}$
with the parameter set given in Table~\ref{initial}. 
In this case, we can find the minimal of the potential at around 10 TeV, and thus the $B-L$ gauge symmetry is broken. 

After the $B-L$ symmetry breaking, i.e., $\varphi$ gets the non-zero VEV, 
the mass terms for $\Phi$ and $\Delta$ fields are effectively generated through the $\lambda_7$ and $\lambda_8$ terms, respectively. 
Furthermore, the $\lambda_0$ term gives the trilinear $\mu$ term as mentioned in the beginning of Sec.~\ref{sec:model}. 
We thus rewrite the dimensionless coupling constants as follows: 
\begin{align}
\mu_\Phi^2    =  \frac{\lambda_7}{2}   v_\varphi^2, \quad 
\mu_\Delta^2  =  \frac{\lambda_8}{2}   v_\varphi^2, \quad 
\mu = \frac{\lambda_0}{\sqrt{2}} v_\varphi.  \label{mus}
\end{align}
In order to realize the spontaneous electroweak symmetry breaking which dominantly happens 
by the doublet Higgs VEV because of the constraint from the electroweak precision data (see the subsection~II-D),  
a negative value of $\mu_\Phi^2$ is required, which indicates necessity of a negative value of the $\lambda_7$ parameter. 
In Sec.~\ref{sec:RGE}, we numerically demonstrate that the $\lambda_7$ parameter gets a negative value at the $B-L$ breaking scale. 
Therefore, the electroweak symmetry breaking is successfully triggered by the radiative breaking of the $U(1)_{B-L}$ symmetry. 

At the scale below $v_\varphi$, the Higgs potential is then effectively described as follows:
\begin{align}
V(\tilde{\varphi},\Phi,\Delta)
&= \mu_\Phi^2 \Phi^\dagger\Phi + \mu_\Delta^2 \text{Tr}(\Delta^\dagger \Delta ) 
+ \mu\left(\Phi^Ti\tau_2 \Delta^\dag\Phi+{\rm h.c.}\right)\notag\\
&+\lambda_1 (\Phi^\dag\Phi)^2 
  + \lambda_2 [{\rm Tr}(\Delta^\dag\Delta)]^2   + \lambda_3 {\rm Tr}[(\Delta^\dag\Delta)^2]
  + \lambda_4 (\Phi^\dag\Phi){\rm Tr}(\Delta^\dag\Delta)   +\lambda_5 \Phi^\dagger \Delta \Delta^\dagger \Phi  \nn\\
 &+\frac{\lambda_6}{4} \tilde{\varphi}^4
  +\frac{\lambda_7}{2}  (\Phi^\dag\Phi)\left( \tilde{\varphi}^2 -v_\varphi^2 \right)  
  +\frac{\lambda_8}{2} {\rm Tr}(\Delta^\dag\Delta)\left( \tilde{\varphi}^2 -v_\varphi^2 \right) \nn\\
  &+\frac{\lambda_0}{\sqrt{2}} \Phi^Ti\tau_2 \Delta^\dag\Phi \left( \tilde{\varphi} -v_\varphi\right)+\text{h.c.} 
 \label{HP2}
\end{align}
The stationary conditions for $\phi_R^{}$ and $\Delta_R$  are given as
\begin{align}
\left.\frac{\partial V}{\partial \phi_R^{}}\right|_0 = 0,\quad
\left.\frac{\partial V}{\partial \Delta_R^{}}\right|_0 = 0. 
\end{align}
They give the following equations
\begin{align}
&\mu_\Phi^2 v_\phi  +\frac{1}{2}v_\phi\left[2\lambda_1 v_\phi^2+(\lambda_4+\lambda_5)v_\Delta^2    \right] -\sqrt{2}\mu v_\phi v_\Delta=0, \\ 
&\mu_\Delta^2v_\Delta^{}    + \frac{1}{2}v_\Delta \left[ 2(\lambda_2 + \lambda_3)v_\Delta^2+(\lambda_4+\lambda_5) v_\phi^2    \right]-\frac{\mu}{\sqrt{2}} v_\phi^2=0. 
\end{align}
From the above two equations, the two VEVs are given under $v_\Delta \ll v_\phi$ as
\begin{align}
v_\phi^2 &\simeq -\frac{\mu_\Phi^2}{\lambda_1}, \quad 
 v_\Delta \simeq \frac{\mu v_\phi^2}{\sqrt2 \left[\mu_\Delta^2+\frac12 \left(\lambda_4+\lambda_5\right)v^2\right]}.  
 \label{eq:tad3}
 \end{align}

In the following, we calculate the mass formulae of the physical scalar states from the potential given in Eq.~(\ref{HP2}). 
In our model, there are one pair of doubly-charged, one pair of singly-charged, one CP-odd and three CP-even physical scalar states. 
The squared masses of the doubly-charged ($H^{\pm\pm}$), the singly-charged ($H^{\pm}$) and the CP-odd ($A$) scalar bosons 
which almost consist of the component fields of $\Delta$ are given by 
\begin{align}
m_{H^{\pm\pm}}^2 &= M_\Delta^2  -\frac{\lambda_5}{2}v_\phi^2  -v_\Delta^2 \lambda_3,  \\
m_{H^{\pm}}^2 &= \left(M_\Delta^2  -\frac{\lambda_5}{4}v_\phi^2\right)\left(1 + \frac{2v_\Delta^2}{v_\phi^2} \right), \\
m_A^2 &= M_\Delta^2\left(1 + \frac{4v_\Delta^2}{v_\phi^2} + \frac{v_\Delta^2}{v_\varphi^2}\right), 
\end{align}
where 
\begin{align}
M_\Delta^2 \equiv \frac{\mu v_\phi^2 }{\sqrt{2}v_\Delta}.  
%\sim \frac{\lambda_8}{2}v_\varphi^2+\frac{\lambda_4+\lambda_5}{2}v_\phi^2. 
\label{eq:mdelta}
\end{align}
The mass term for the CP-even scalar states is obtained from the second derivatives of the CP-even scalar states. 
We note that only for the $\varphi_R^{2}$ term, the dominant contribution comes from the one-loop effect as shown in Eq.~(\ref{mphisq}).
We then obtain the mass term as 
\begin{align}
V_{\text{CP-even}}^{\text{mass}} &=  \frac{1}{2}(\phi_R^{},\Delta_R^{},\varphi_R^{})M_R^2
\begin{pmatrix}
\phi_R^{}\\
\Delta_R^{}\\
\varphi_R^{}
\end{pmatrix}
=
\frac{1}{2}(H_1,H_2,H_3)\, \text{diag}(m_{H_1}^2,m_{H_2}^2,m_{H_3}^2)\,
\begin{pmatrix}
H_1\\
H_2\\
H_3
\end{pmatrix}, \label{massmat}
\end{align}
where each of the matrix elements is given by 
\begin{align}
(M_R^2)_{11} &= 2\lambda_1 v_\phi^2, \label{m11}\\
(M_R^2)_{22} &= M_\Delta^2 + 2v_\Delta^2(\lambda_2 + \lambda_3), \\
(M_R^2)_{33} &=-4\lambda_6 v_\varphi^2,   \\
(M_R^2)_{12} &= \frac{v_\Delta}{v_\phi}[ v_\phi^2 (\lambda_4 + \lambda_5)-2M_\Delta^2], \\ 
(M_R^2)_{13} &= -2\lambda_1\frac{v_\phi^3}{v_\varphi}
-\left(\lambda_4+\lambda_5\right)\frac{v_\phi v_\Delta^2}{v_\varphi}+2\frac{M_\Delta^2v_\Delta^2}{v_\phi v_\varphi}, \\
(M_R^2)_{23} &=  -2\left(\lambda_2+\lambda_3\right)\frac{v_\Delta^3}{v_\varphi}
-\left(\lambda_4+\lambda_5\right)\frac{v_\Delta v_\phi^2}{v_\varphi}+\frac{M_\Delta^2v_\Delta}{v_\varphi}. \label{m23}
\end{align}
In Eq.~(\ref{massmat}), $H_1$, $H_2$ and $H_3$ are the mass eigenstates, and $m_{H_1}^{}$, $m_{H_2}^{}$ and $m_{H_3}^{}$ 
($m_{H_1}^{} \leq m_{H_2}^{} \leq m_{H_3}^{}$)
are the corresponding mass eigenvalues. 
The mass eigenstates are related to the weak eigenstates by an orthogonal matrix $R$ as 
\begin{align}
\begin{pmatrix}
\phi_R \\
\Delta_R \\
\varphi_R
\end{pmatrix} = R
\begin{pmatrix}
H_1 \\
H_2 \\
H_3
\end{pmatrix}, 
\end{align}
where $R$ can be described by three mixing angles. 
Therefore, the six independent matrix elements given in Eqs.~(\ref{m11})-(\ref{m23}) are described by the 
three masses and three mixing angles.   
Because of the discovery of the Higgs boson at the LHC, one of the CP-even Higgs bosons must be identified as the discovered one. 
We can, for example, regard $H_1$ as the Higgs boson with mass of 125 GeV, i.e., $m_{H_1}^{}=125$ GeV. 

It is important to mention here that there appears a characteristic relationship among the masses of $H^{\pm\pm}$,   
$H^{\pm}$ and $A$ under $v_\Delta \ll v$ as~\cite{Chun,AKY,Melfo,triplet1} 
\begin{align}
m_{H^{\pm\pm}}^2 - m_{H^\pm}^2 = 
m_{H^{\pm}}^2 - m_{A}^2 = -\frac{\lambda_5}{4}v^2 + {\cal O}(v_\Delta^2/v^2). 
\end{align}
Therefore, these three mass parameters are determined by two input parameters, e.g., $M_\Delta^2$ and $\lambda_5$ if we neglect 
the ${\cal O}(v_\Delta^2/v^2)$ correction. 
In addition, the sign of the $\lambda_5$ parameter determines the pattern of the mass hierarchy, namely, 
if $\lambda_{5}>0$ ($\lambda_{5} < 0$), $m_{H^{\pm\pm}}^{}< m_{H^\pm}^{}<m_A^{}$ ($m_{H^{\pm\pm}}^{} > m_{H^\pm}^{} > m_A^{}$). 
The phenomenology of these Higgs bosons can be drastically different depending on the pattern of the mass hierarchy. 
For example, the decay pattern of $H^{\pm\pm}$, which is quite important to test the HTM, 
strongly depends on the mass spectrum. 
If we consider the case of $\lambda_{5} < 0$, $H^{\pm\pm}$ can mainly decay into the singly-charged Higgs boson $H^{\pm}$ and  
the $W$ boson. Collider signatures for this case at the LHC have been simulated in Ref.~\cite{AKY}. 
On the other hand, if we consider the case of $\lambda_{5} > 0$, $H^{\pm\pm}$ can mainly decay into the same sign dilepton or 
the same sign diboson depending on the magnitude of the Yukawa coupling $h_{ij}$ defined in Eq.~(\ref{Yukawa}) and the triplet VEV $v_\Delta^{}$. 
In the minimal HTM, 
in the case of $v_\Delta \lesssim 0.1$ ($v_\Delta \gtrsim 0.1$) MeV, 
$H^{\pm\pm}$ mainly decay into the dilepton~\cite{dilep1,dilep2,dilep3} (diboson~\cite{Nomura,Yagyu,WW}). 
In this case, the decay of the singly-charged Higgs boson $H^\pm$ into $H^{\pm\pm}$ and the $W$ boson can increase the number of 
events rate for $H^{\pm\pm}$~\cite{Akeroyd}. 
In our scenario with the CCI, the mass spectrum can be predicted by using the one-loop RGEs as it will be discussed in Sec.~\ref{sec:RGE}.

\subsection{Lepton sector}

The Yukawa Lagrangian for the lepton sector is given by 
\begin{align}
{\cal L}_Y&=
y_\ell^{i} \overline{L_L^i} \Phi e_R^i 
+ y_N^{i} \varphi \overline{ N_R^{ci}} N_R^i
+ y_D^{ij} \overline{L_L^i} (i\tau_2)\Phi^* N_R^j 
+ y_\Delta^{ij} \overline{L_L^{ic}} \Delta L_L^j  +{\rm h.c.} \label{Yukawa}, 
\end{align}
where the first term is the same as the Yukawa interaction for leptons in the SM. 
The second and third terms respectively give the Majorana masses for $N_R^i$ and the Dirac masses for left- and right-handed neutrinos. 
Therefore, the type-I seesaw mechanism~\cite{typeI} is realized by these two terms. 
Finally, the last term in Eq.~(\ref{Yukawa}) also gives Majorana masses for left-handed neutrinos via the type-II seesaw mechanism~\cite{typeII}. 
As a result, in our model, 
the neutrino mass generation corresponds to the hybrid 
scenario based on the type-I and type-II seesaw mechanisms~\cite{hybrid} expressed as 
\begin{align}
& (m_\nu)_{ij} = (m_\nu^{\text{I}})_{ij} +  (m_\nu^{\text{II}})_{ij}. 
\end{align}
The type-I and type-II contributions are respectively expressed as 
\begin{align}
(m_\nu^{\text{I}})_{ij}  =  v_\phi^2 y_D^{ik} (M_N^{k})^{-1}y_D^{jk}, \quad 
(m_\nu^{\text{II}})_{ij} = \sqrt{2}y_\Delta^{ij} v_\Delta^{},
\end{align}
where $M_N^k = \sqrt2 y_N^k v_\varphi$.

Regarding to the type-II contribution $m_\nu^{\text{II}}$, 
the magnitude of the Yukawa coupling $y_\Delta^{}$ is constrained by LFV data. 
In our model, there are two types of the LFV processes, namely the tree level
$\ell_i\to\ell_j\ell_k\ell_\ell$ type processes and 
the one-loop $\ell_i\to\ell_j\gamma$ type processes.  
The analytic expressions for the branching fractions of these LFV processes are obtained by 
\begin{align}
&{\cal B} (\ell_i^\pm \to \ell_j^\mp \ell_k^\pm \ell_\ell^\pm) =\frac{1}{8 G_F^2}\frac{1}{m^4_{H^{\pm\pm}}} \left|y_\Delta^{k\ell *}\, y_\Delta^{ji} \right|^2(1+\delta_{k\ell}) , \\
&{\cal B}(\ell_i^{\pm} \to \ell_j^{\pm}\gamma) 
= \frac{\alpha_{\text{em}}}{192 G_F^2}\left(\frac{8}{ m^2_{H^{\pm\pm}}}+\frac{1}{ m^2_{H^{\pm}}}\right)^2\left|\sum_{a=1}^3y^{aj*}_\Delta\, y_\Delta^{ai}\right|^2. 
\end{align}
By comparing the measured branching fractions of the LFV processes 
and those model predictions, we obtain the following constraints on the 
combination of $y_\Delta^{ij}$ couplings as~\cite{Herrero-Garcia:2014hfa}
\begin{align}
\begin{array}{lc}
|y_\Delta^{12} y_\Delta^{11*}|< 2.3\times 10^{-5}\times \left(\frac{{m_{H^{\pm\pm}}^{}}}{{\rm TeV}}\right)^2, &\text{from}~~ \mu^-\to e^+e^-e^-, \\
 |y_\Delta^{13} y_\Delta^{11*}|< 0.009\times \left(\frac{{m_{H^{\pm\pm}}^{}}}{{\rm TeV}}\right)^2, &  \text{from}~~ \mu^-\to e^+e^-e^- \\
 |y_\Delta^{13} y_\Delta^{12*}|< 0.005\times \left(\frac{{m_{H^{\pm\pm}}^{}}}{{\rm TeV}}\right)^2, &\text{from}~~ \tau^-\to e^+e^-\mu^-, \\
 |y_\Delta^{13} y_\Delta^{22*}|< 0.007\times \left(\frac{{m_{H^{\pm\pm}}^{}}}{{\rm TeV}}\right)^2, &\text{from}~~ \tau^-\to e^+\mu^-\mu^-, \\
 |y_\Delta^{23} y_\Delta^{11*}|< 0.007\times \left(\frac{{m_{H^{\pm\pm}}^{}}}{{\rm TeV}}\right)^2, &\text{from}~~ \tau^-\to \mu^+e^-e^-,   \\
 |y_\Delta^{23} y_\Delta^{12*}|< 0.007\times \left(\frac{{m_{H^{\pm\pm}}^{}}}{{\rm TeV}}\right)^2, &\text{from}~~ \tau^-\to \mu^+e^-\mu^-,  \\
 |y_\Delta^{23} y_\Delta^{22*}|< 0.008\times \left(\frac{{m_{\Delta^{\pm\pm}}^{}}}{{\rm TeV}}\right)^2, &\text{from}~~ \tau^-\to \mu^+\mu^-\mu^-,
	\label{eq:constraint_2}
\end{array}
\end{align}
and
\begin{align}
\begin{array}{lc}
 |y_\Delta^{11*}y_\Delta^{12}  + y_\Delta^{21*} y_\Delta^{22}  + y_\Delta^{31*} y_\Delta^{32}|^2 < 2.0\times 10^{-6}
\times \frac{m_{H^{\pm}}^4 m_{H^{\pm\pm}}^4}{(8m_{H^{\pm}}^2+m_{H^{\pm\pm}}^2)^2} \frac{1}{({\rm TeV})^4}, & \text{from}~~\mu\to e\gamma ,\\
 |y_\Delta^{11*} y_\Delta^{13}  + y_\Delta^{21*} y_\Delta^{23}  + y_\Delta^{31*} y_\Delta^{33}|^2 < 0.12\times 
 \frac{m_{H^{\pm}}^4 m_{H^{\pm\pm}}^4}{(8m_{H^{\pm}}^2+m_{H^{\pm\pm}}^2)^2} \frac{1}{({\rm TeV})^4},& \text{from}~~
\tau\to e\gamma ,\\ 
 |y_\Delta^{12*} y_\Delta^{13}  + y_\Delta^{22*} y_\Delta^{23}  + y_\Delta^{32*} y_\Delta^{33}|^2 < 0.16\times 
 \frac{m_{H^{\pm}}^4 m_{H^{\pm\pm}}^4}{(8m_{H^{\pm}}^2+m_{H^{\pm\pm}}^2)^2} \frac{1}{({\rm TeV})^4},& \text{from}~~\tau\to \mu \gamma .  
	\label{eq:constraint_3}
\end{array}
\end{align}
%}

\subsection{Kinetic term}\label{sec:kin}

The kinetic terms for the scalar fields are given by 
\begin{align}
{\cal L}_{\text{kin}} = (D_\mu \Phi)^\dagger(D^\mu \Phi) 
+ \text{Tr}(D_\mu\Delta)^\dagger(D^\mu\Delta) 
+ (D_\mu \varphi)^\dagger(D^\mu \varphi), \label{kin}
\end{align}
where the covariant derivatives are expressed as 
\begin{align}
D_\mu \Phi   &= \left(\partial_\mu - i\frac{g_2}{2}\tau^a A_\mu^a -i\frac{g_1}{2}B_\mu -i\frac{g_{\text{mix}}}{2}C_\mu  \right)\Phi, \label{kin2}\\
D_\mu \Delta &= \partial_\mu\Delta - i\frac{g_2}{2}[\tau^aA_\mu^a,\Delta] -ig_1B_\mu \Delta  
-i\left(g_{\text{mix}}+2 g_{B-L}\right)C_\mu \Delta, \label{kin3}\\
D_\mu \varphi&= \left(\partial_\mu - 2ig_{B-L}^{}C_\mu  \right)\varphi, \label{kin4}
\end{align}
with ($g_2$, $g_1$, $g_{B-L}^{}$) and ($A_\mu^a$, $B_\mu$, $C_{\mu}$) being the ($SU(2)_L$,~$U(1)_Y$,~$U(1)_{B-L}$) gauge coupling constants and corresponding gauge fields, respectively. 
The $g_{\text{mix}}$ coupling is defined so as to be absent the kinetic mixing between the $U(1)_Y$ and $U(1)_{B-L}$ gauge bosons. 
After the $B-L$ and electroweak symmetry breaking, 
the mass of the $W$ boson is given as 
\begin{align}
m_W^2 = \frac{g_2^2}{4}v^2, 
\end{align}
where $v^2=v_\phi^2 + 2v_\Delta^2=(\sqrt{2}G_F)^{-1}\simeq (246$ GeV)$^2$. 
For the neutral gauge bosons, 
the photon state $A_\mu$ is obtained by the linear combination of $A_\mu^3$ and $B_\mu$ fields as in the SM:
\begin{align}
A_\mu = B_\mu \cos\theta_W^{} + A_\mu^3\sin\theta_W^{}, \quad 
\tilde{Z}_\mu = -B_\mu \sin\theta_W^{} + A_\mu^3\cos\theta_W^{}, 
\end{align}
where $\tilde{Z}_\mu$ is the orthogonal state for $A_\mu$ which can be mixed with the $C_\mu$ field. 
The mass matrix for the massive neutral gauge bosons $M_{\text{gauge}}^2$ in the basis of ($\tilde{Z}_\mu$,$C_\mu$) is given by 
 \begin{align}
&M_{\text{gauge}}^2 =  \notag\\
&\hspace{-3mm}\begin{pmatrix}
 \frac{g_Z^2}{4}(v^2 + 2v_\Delta^2) & -\frac{g_Z^{}}{4}[g_{\text{mix}}v^2 + 2(g_{\text{mix}}+4g_{B-L}^{})v_\Delta^2] \\
 -\frac{g_Z^{}}{4}[g_{\text{mix}}v^2 + 2(g_{\text{mix}}+4g_{B-L}^{})v_\Delta^2] & 4g_{B-L}^2(v_\varphi^2 + v_\Delta^2) + \frac{g_{\text{mix}}^2}{4}(v^2 + 2v_\Delta^2) + 4g_{\text{mix}}g_{B-L}^{}v_\Delta^2
 \end{pmatrix}, 
 \end{align}
where $g_Z^{}=g_2/\cos\theta_W^{}$. 
The mass eigenstates for the massive neutral gauge bosons are defined by $Z_\mu$ and $Z_\mu'$ via an $SO(2)$ transformation. 
Under $v_\varphi \gg v,~v_\Delta^{}$, the masses of $Z_\mu$ and $Z_\mu'$ are given by 
\begin{align}
m_Z^2&=\frac{g_Z^2}{4}(v^2 + 2v_\Delta^2)+{\cal O}\left(\frac{v^4}{v_\varphi^2}, \frac{v_\Delta^4}{v_\varphi^2}, \frac{v^2 v_\Delta^2}{v_\varphi^2}\right), \\
m_{Z'}^2&= 4g_{B-L}^2 v_\varphi^2 +\frac{g_{\text{mix}}^2}{4}v^2 +\left(g_{\text{mix}}+2g_{B-L}\right)^2v_\Delta^2
+{\cal O}\left(\frac{v^4}{v_\varphi^2}, \frac{v_\Delta^4}{v_\varphi^2}, \frac{v^2 v_\Delta^2}{v_\varphi^2}\right). 
\end{align}
The electroweak rho parameter $\rho$ deviates from unity at the tree level:
\begin{align}
\rho \equiv \frac{m_W^2}{m_Z^2 \cos^2\theta_W} = \frac{v^2}{v^2 + 2v_\Delta^2}+{\cal O}\left(\frac{v^4}{v_\varphi^2}, \frac{v_\Delta^4}{v_\varphi^2}, \frac{v^2 v_\Delta^2}{v_\varphi^2}\right). 
\end{align}
From the experimental value of the $\rho$ parameter, i.e.,
$\rho_{\text{exp}}=1.0004^{+0.0003}_{-0.0004}$~\cite{PDG}, 
$v_\Delta^{}$ is constrained to be less than a few GeV.

\section{Numerical results} \label{sec:RGE}

\begin{center}
\begin{table}[t]
\begin{tabular}{c|c|c|c|c|c|c||c|c|c||c|c}\hline\hline
\multicolumn{7}{c||}{Scalar couplings } &
\multicolumn{3}{c||}{Yukawa couplings }&
\multicolumn{2}{c}{Gauge couplings} \\\hline
$~~\lambda_1~~$&$~~\lambda_2~~$&$~~\lambda_3~~$&$~~\lambda_4~~$&$~~\lambda_6~~$&$~~\lambda_7~~$&$~~\lambda_8$~~
&$~~y_N^{}~~$&$y_D^{}$&$y_\Delta^{}$
&$~~g_{B-L}^{}~~$&$g_{\text{mix}}^{}$ \\\hline
$0.02$&$0.8$&$0.1$&$0.1$&$0.0715$&$-0.00694$&$0.41$
&$0.104$&$1.35\times 10^{-6}$&$0.257$
&$0.233$&$-0.794$ \\\hline \hline
\end{tabular}
\caption{Initial values of the running coupling constants at the Planck scale. }
\label{initial}
\end{table}
\end{center}

We numerically solve the RGEs to determine the values of
the scalar quartic couplings at the low energy scale and to obtain predictions of the low energy observables such as 
the mass spectrum of the scalar bosons. 
The full set of analytic formulae for the beta functions of all the
gauge, Yukawa and scalar couplings are given in Appendix.  

As we mentioned in Sec.~\ref{sec:model}, 
the global symmetry given in Eq.~(\ref{global}) is expected to be restored in the Higgs potential, by which 
$\lambda_0$ and $\lambda_5$ terms are forbidden at a high energy scale. 
We thus set the initial values of $\lambda_0$ and $\lambda_5$ to be zero at the Planck scale $M_{\text{pl}}=2.4\times 10^{18}$ GeV:
\begin{eqnarray}
\lambda_0\left(M_{\text{pl}}\right) =\lambda_5\left(M_{\text{pl}}\right)=0. 
\end{eqnarray}
Initial values for all the other coupling constants 
should be taken to realize the $B-L$ and electroweak symmetry breaking, to satisfy Eq.~(\ref{eq:tad3}), and to reproduce the 
125 GeV Higgs boson mass and the correct order of the neutrino masses, i.e., ${\cal O }(0.1)$ eV.  
We find a set of such initial value
%\footnote{For the gauge and Yukawa couplings, we give 
%initial values at $v_\varphi$ and $m_t$, respectively. } 
in Table~\ref{initial}, where 
all the Yukawa coupling matrices $y_N^{}$, $y_D^{}$ and $y_\Delta^{}$ are 
assumed to be proportional to the $3\times 3$ identity matrix for simplicity.
With these initial values of gauge and Yukawa couplings, we can determine the values of these couplings
at low energies, because the running of these couplings are closed by themselves at the one-loop level.  
At the $B-L$ scale, we obtain $y_N^{}(v_\varphi)\simeq 0.10$, $y_D^{}(v_\varphi)\simeq 1.3\times 10^{-6}$ and $y_\Delta^{}(v_\varphi)\simeq 0.17$ which 
gives $m_\nu^{\text{I}}\simeq 0.080$ eV and $m_\nu^{\text{II}} \simeq 1.2\times 10^{-3}$ eV. 

\begin{figure}[t]
\begin{center}
\includegraphics[scale=0.6]{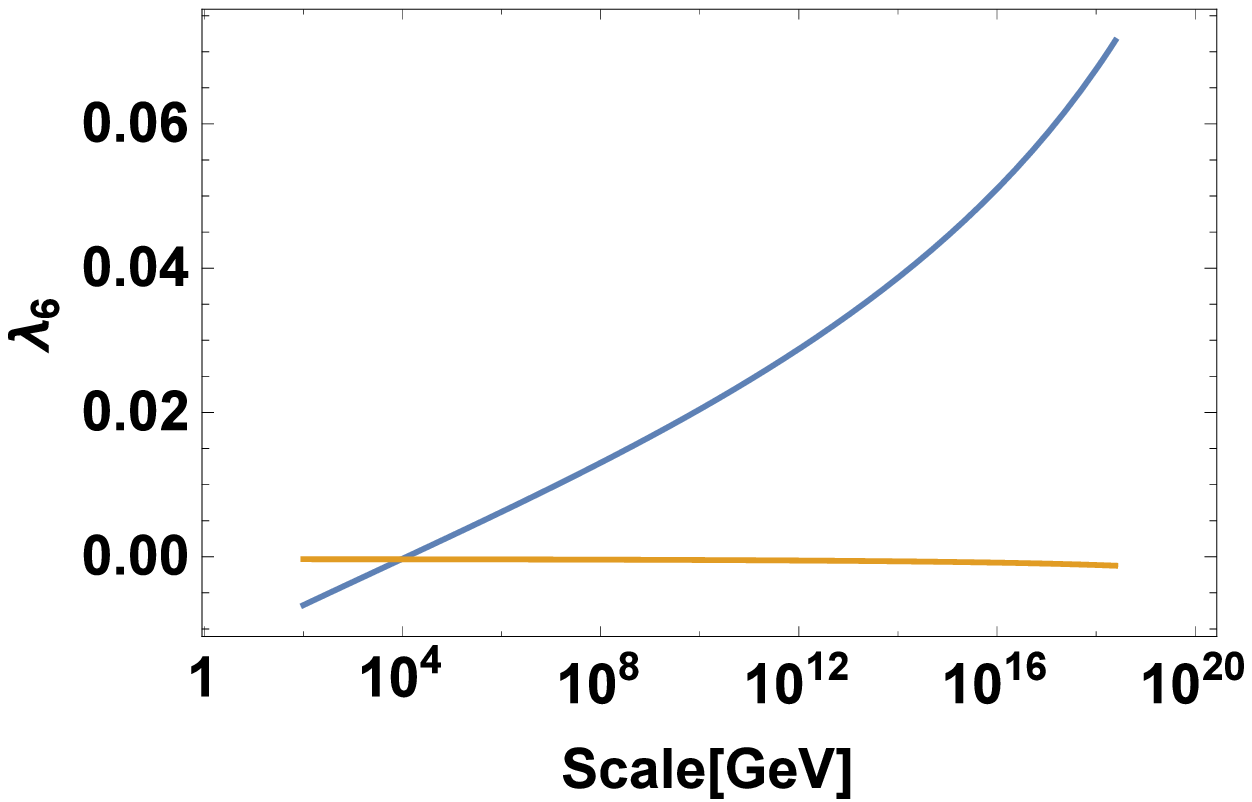}\hspace{5mm}
\includegraphics[scale=0.6]{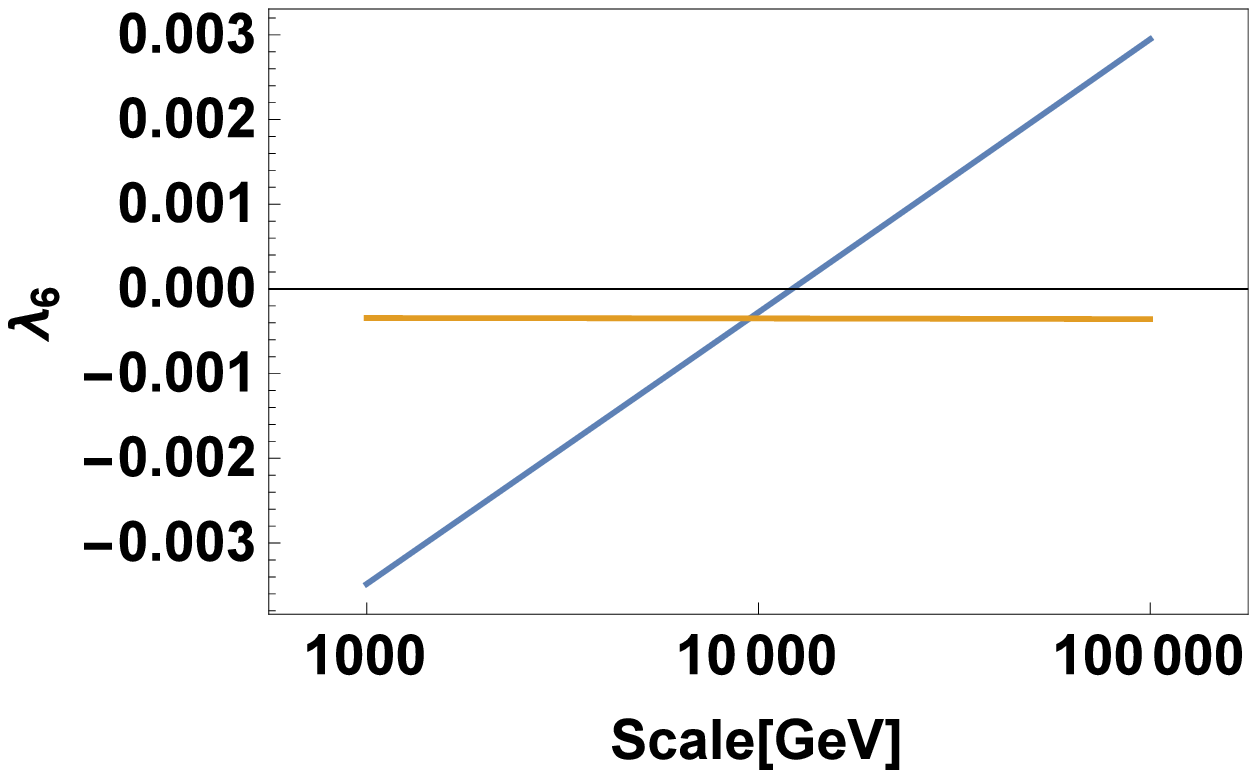}
\caption{The scale dependence of $\lambda_6$.  
The right panel focuses on the scale from $10^3$ to $10^5$ GeV. 
We use the initial values of the coupling constants given in Table~\ref{initial}. 
The orange curve represents the right-hand side of Eq.~(\ref{eq:tad1}).}
\label{fig:lam6}
\end{center}
\end{figure}

First, we discuss the spontaneous breakdown of the $U(1)_{B-L}$ gauge symmetry. 
In Fig.~\ref{fig:lam6}, we show the scale dependence of the $\lambda_6$ parameter. 
The horizontal line denotes the right hand side of Eq.~(\ref{eq:tad1}). 
The intersection point of two curves determines the $B-L$ breaking scale, and in this case, 
it is determined to be 9.48 TeV, i.e., $v_\varphi=9.48$ TeV as it is also seen in Fig.~\ref{fig:potential}

Second, we discuss the electroweak symmetry breaking. 
This can be confirmed by checking that $\lambda_7$ is given to be a negative value at the $B-L$ breaking scale. 
In Fig.~\ref{fig:lam6}, we show the RGE running of the $\lambda_7$ parameter. 
We can see that the $\lambda_7$ parameter have a negative value at around 10 TeV, so that the spontaneous electroweak 
symmetry breaking is successfully realized. 

\begin{figure}[t]
\begin{center}
\includegraphics[scale=0.6]{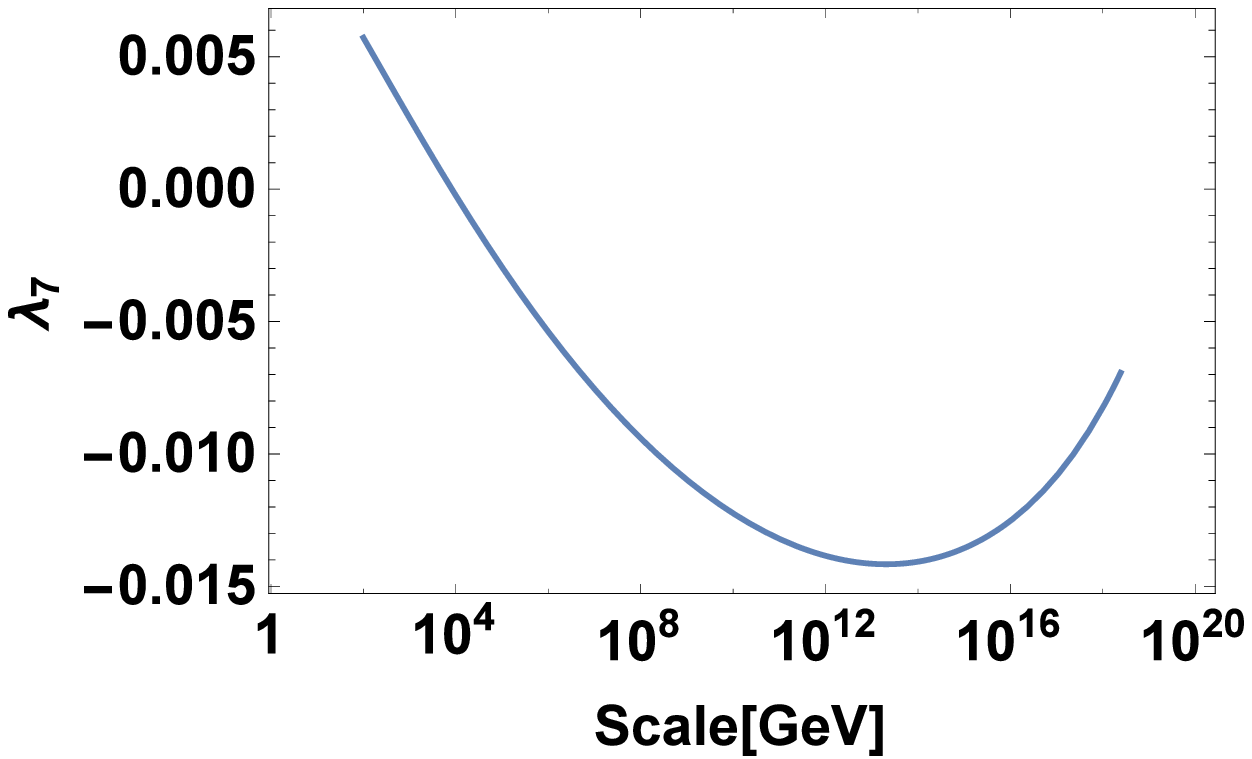}\hspace{5mm}
\includegraphics[scale=0.6]{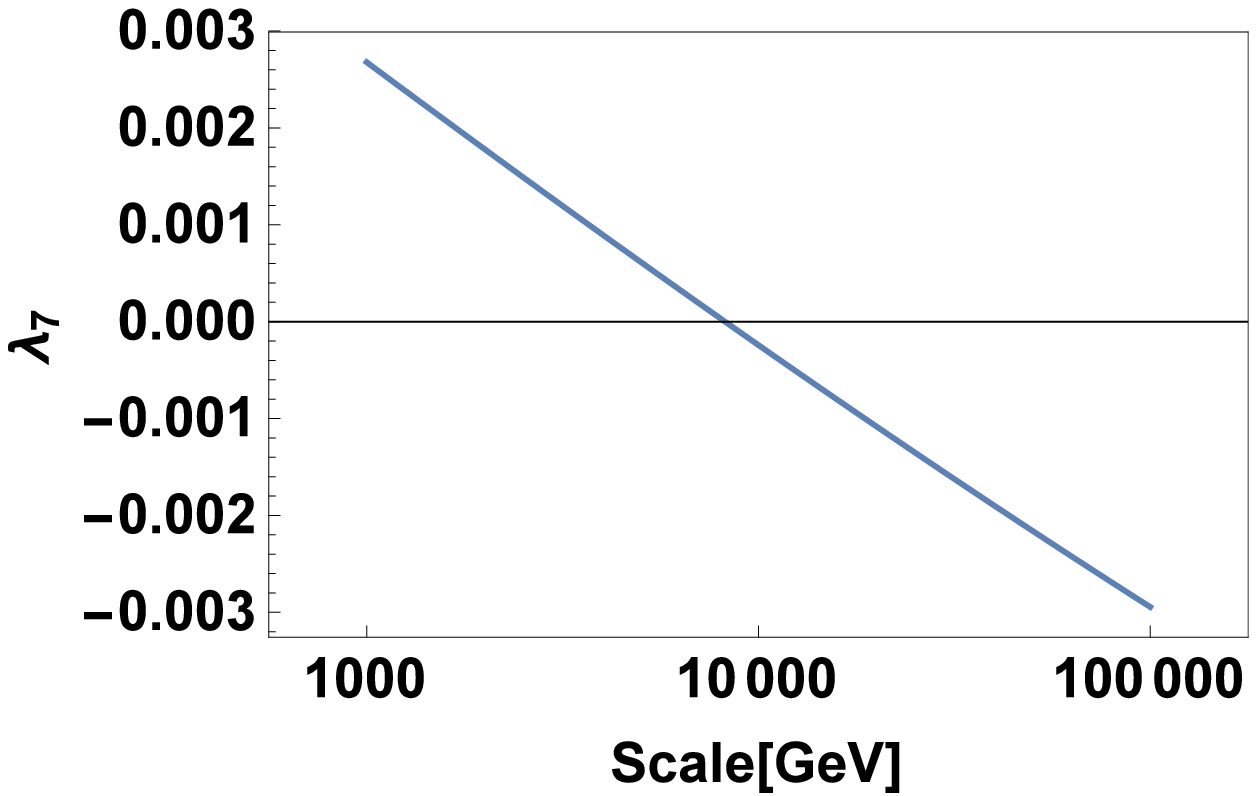}
\caption{The scale dependence of $\lambda_7$.  
The right panel focuses on the scale from $10^3$ to $10^5$ GeV. 
We use the initial values of the coupling constants given in Table~\ref{initial}. }
\label{fig:lam7}
\end{center}
\end{figure}

\begin{figure}[t]
\begin{center}
\includegraphics[scale=0.6]{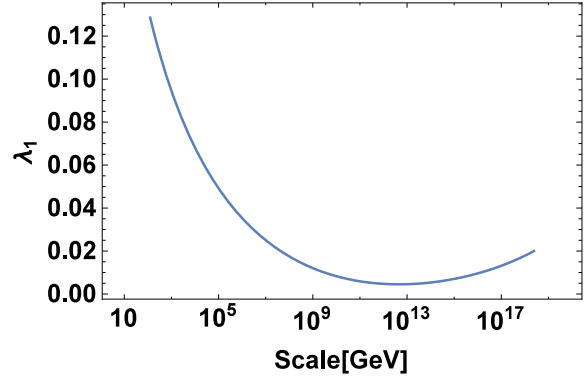}
\caption{The scale dependence of $\lambda_1$. 
We use the initial values of the coupling constants given in Table~\ref{initial}. }
\label{fig:lam1}
\end{center}
\end{figure}

\begin{figure}[t]
\begin{center}
\includegraphics[scale=0.6]{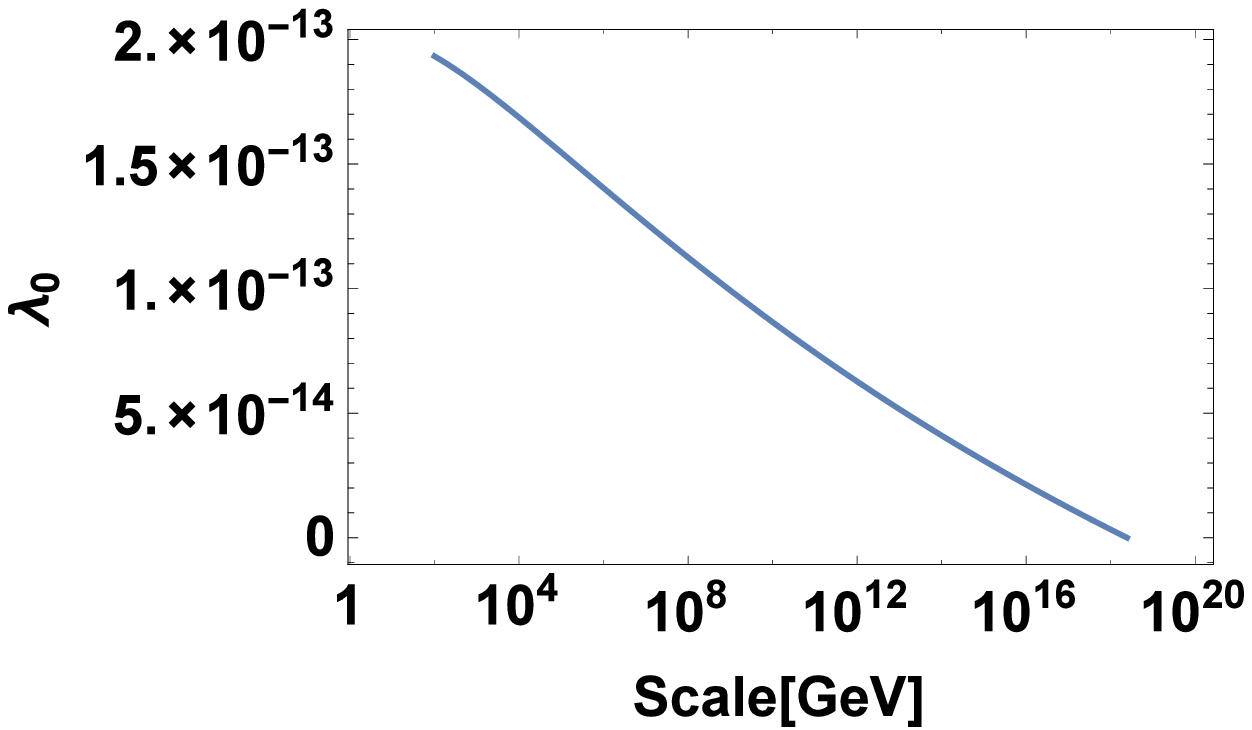}\hspace{5mm}
\includegraphics[scale=0.6]{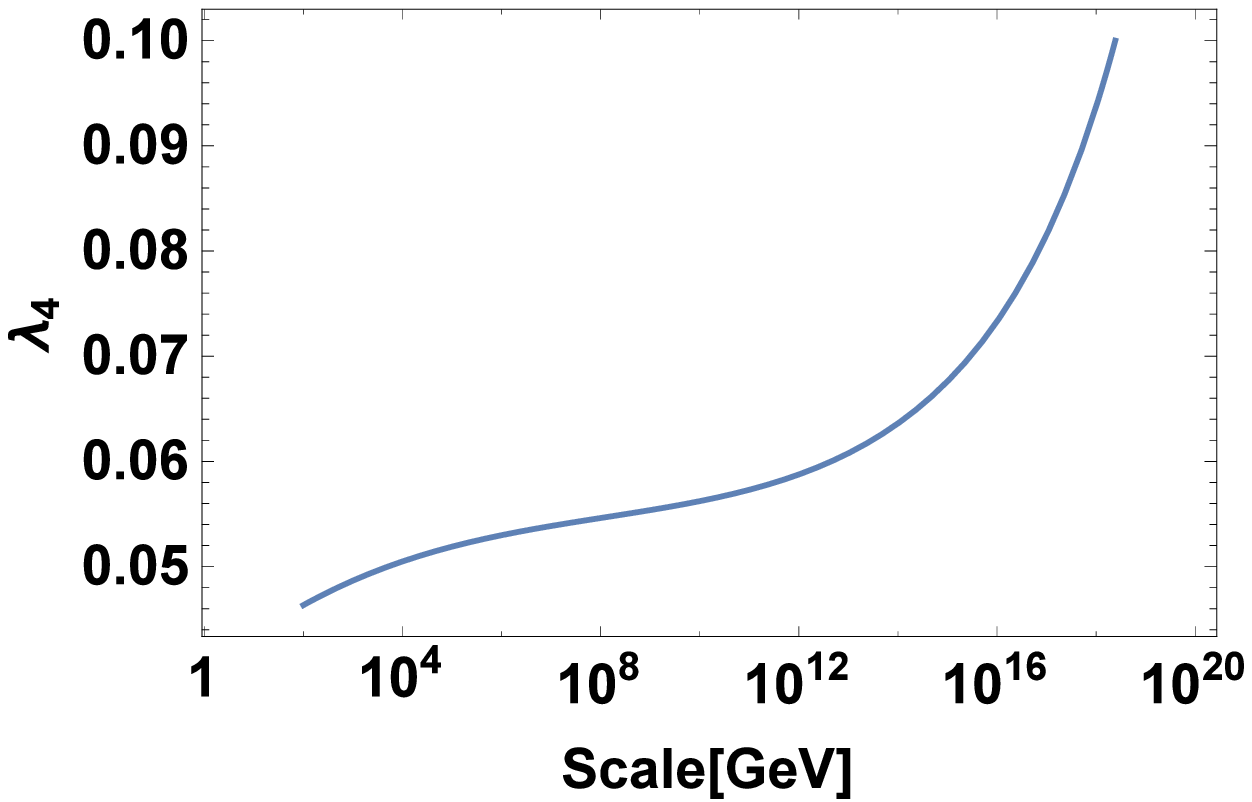} \\ \vspace{10mm}
\includegraphics[scale=0.6]{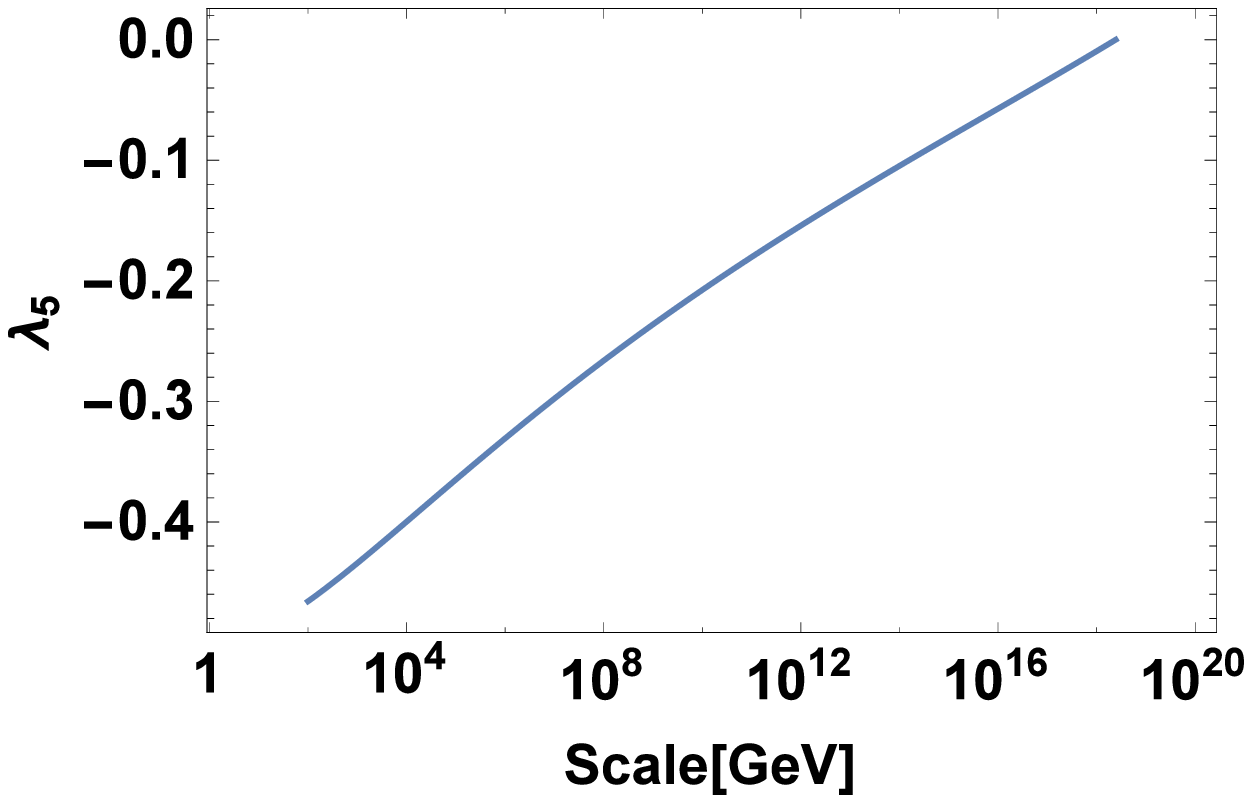}\hspace{5mm}
\includegraphics[scale=0.6]{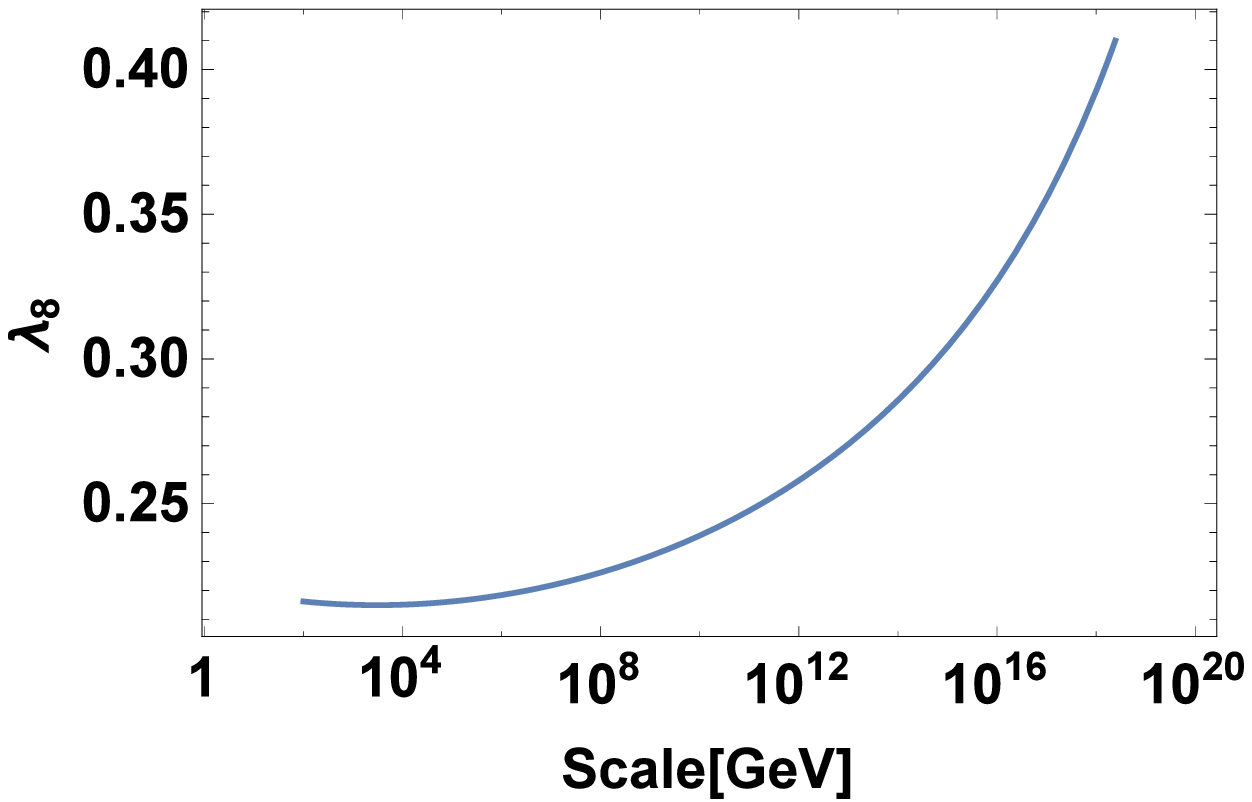}
\caption{Scale dependence of $\lambda_0$, $\lambda_4$, $\lambda_5$ and $\lambda_8$.  
We use the initial values of the coupling constants given in Table~\ref{initial}. 
}
\label{fig:lams}
\end{center}
\end{figure}

Third, we show the running of $\lambda_1$ in Fig.~\ref{fig:lam1}. 
At the scale around 100 GeV, we can see that $\lambda_1\simeq 0.12$ is 
given which reproduces the Higgs boson mass to be about 125 GeV.  

Fourth, the RGE running of $\lambda_0$, $\lambda_4$, $\lambda_5$ and $\lambda_8$ is shown in Fig.~\ref{fig:lams}.
The other couplings $\lambda_2$ and $\lambda_3$ do not give an important impact on the 
low energy observables, so that we do not show their running's.
With determined these $\lambda$ couplings at the $B-L$ breaking scale, we obtain the predictions for the following quantities:
\begin{align}
&v_\Delta^{} =5.03\times 10^{-12}~\text{GeV}, \quad
\mu=1.13\times10^{-9} \ \text{GeV}, \quad 
M_\Delta^2 = (3.10\times  10^3)^2~\text{GeV}^2,  \notag\\
&m_{H^{\pm\pm}}^{} =3108~\text{GeV}, \quad
m_{H^{\pm}}^{} =3107~\text{GeV} , \quad
m_A^{} =3106~\text{GeV}, \\
&m_{H_1}^{} =124.9~\text{GeV}, \quad m_{H_2}^{}=355.1~\text{GeV}, \quad m_{H_3}^{}=3106~\text{GeV}, \notag\\ 
& R = \begin{pmatrix}
1 & 0 & 0 \\
0 & 0 & 1 \\
0 & 1 & 0 
\end{pmatrix}  + {\cal O}(10^{-3}).  \label{r}
\end{align}
Notice that the result for $R$ tells us that $H_1$, $H_2$ and $H_3$ are almost 
$\phi_R^{}$, $\varphi_R^{}$ and $\Delta_R^{}$, respectively. 
Around 3 TeV masses for $H^{\pm\pm}$, $H^\pm$, $A$ and $H_3$ are due to $M_\Delta\simeq 3$ TeV which is determined by 
the ratio $\mu/v_\Delta^{}$ as it is seen in  Eq.~(\ref{eq:mdelta}).
Furthermore, the order of the ratio $\mu/v_\Delta^{}$ is determined 
by $\mu_\Delta^2/v^2 \simeq \lambda_8\,v_\varphi^2/v^2 = {\cal O}(10^2) $ from Eqs.~(\ref{eq:tad3}) and (\ref{mus}) with $\lambda_8= {\cal O}(0.1)$. 
Thus, the magnitude of $M_\Delta$ can be roughly explained by $10\times v$. 

\begin{center}
\begin{table}[t]
\begin{tabular}{c|c|c|c|c|c|c  }\hline\hline  
\multicolumn{7}{c}{Bounds from $\ell \to 3\ell$}   \\\hline
          $~~|y_\Delta^{12}\, y_\Delta^{11*}|~~$&$~~|y_\Delta^{13} \, y_\Delta^{11*}|~~$&$~~|y_\Delta^{13}\, y_\Delta^{12*}|~~$
         &$~~|y_\Delta^{13}\, y_\Delta^{22*}|~~$&$~~|y_\Delta^{23}\, y_\Delta^{11*}|~~$&$~~|y_\Delta^{23}\, y_\Delta^{12*}|~~$&$~~|y_\Delta^{23}\, y_\Delta^{22*}|$~~  \\\hline 
          $2.2\times 10^{-4}$&$0.087$&$0.048$&$0.068$&$0.068$&$0.068$&$0.077$ \\\hline \hline 
\end{tabular} \\ \vspace{5mm}
%%%
\begin{tabular}{c|c|c  }\hline\hline  
\multicolumn{3}{c}{Bounds from $\ell \to \ell'\gamma$}   \\\hline
          $~~\left|\sum_{a=1}^3y_\Delta^{1a*}\, y_\Delta^{2a} \right|^2~~$&$~~\left|\sum_{a=1}^3y_\Delta^{1a*}\, y_\Delta^{3a} \right|^2~~$
&$~~\left|\sum_{a=1}^3y_\Delta^{2a*}\, y_\Delta^{3a} \right|^2$~~  \\\hline 
          $2.4\times 10^{-6}$&0.14&0.18 \\\hline \hline 
\end{tabular}
\caption{Upper limit on the combinations of $y_\Delta^{ij}$ from the LFV data under the predictions given in Eq.~(\ref{r}).  }
\label{Tab:LFV}
\end{table}
\end{center}

Finally, we discuss the constraint from the LFV processes as it was discussed in Sec.~II-C under the prediction given in Eq.~(\ref{r}). 
Since the masses of the doubly- and singly-charged Higgs bosons are 
determined, predictions for the constraints on $y_\Delta^{ij}$ 
couplings are obtained from the LFV data given in Eqs.~(\ref{eq:constraint_2}) and (\ref{eq:constraint_3}). 
In Table~\ref{Tab:LFV}, we list the upper limits on combinations of $y_\Delta^{ij}$ couplings from the $\ell \to 3\ell$ and $\ell \to \ell'\gamma$ types of LFV processes.
%%%
The most severe constraint is obtained for the value of $|y_\Delta^{12}y_\Delta^{11*}|\approx2.2\times10^{-4}$ from the $\mu\to 3e$ process as compared to the other combinations. 
%%%%%%%
\if0
We thus show the allowed parameter space, which is depicted by red points, on the $y_\Delta^{11}$ and $y_\Delta^{12}$ plane in Fig.~\ref{fig:y11-y12} with generic numerical analysis. 
In this figure, we take $\text{Max}(y_\Delta^{ij}) \leq 0.293$ because of the initial value of $y_\Delta^{}(m_t)=0.293$ shown in Table~\ref{initial}. 
%We find that the upper bound on $y_\Delta^{12}$ is given to be about 0.012. 
The black solid line is obtained by solving the upper bound process of $\mu\to3e$; $y_\Delta^{12}=2.3\times10^{-5}\frac{M_{H^{\pm\pm}}^2}{y^{11}_\Delta}$, that provides the origin for giving the most stringent constraint.
We find that the value of $y_\Delta^{12}$ closed to 0.293 tend to be disfavored.
%\textcolor{red}{How about y11?}. 
\begin{figure}[t]
\includegraphics[scale=0.7]{y11-y12.eps}
\caption{Constraint on the parameter space in the $y_\Delta^{11}$ and $y_\Delta^{12}$ plane from the LFV data, where each of dots is allowed. }
\label{fig:y11-y12}
\end{figure}
\fi
%%%%%%%

\section{Conclusion}

We have discussed the extension of the minimal HTM with the CCI and with the gauged $U(1)_{B-L}$ symmetry. 
Tiny neutrino masses are generated by the hybrid mechanism of the type-I and the type-II seesaw. 
In order to determine the 
shape of the Higgs potential at low energies, 
we have prepared the analytic formulae for the one-loop beta functions of all the dimensionless coupling constants. 
We have found the set of the initial values of the parameters at the Planck scale, by which 
the $U(1)_{B-L}$ symmetry is radiatively broken via the Coleman-Weinberg mechanism at a {\cal O}(10) TeV scale. 
The electroweak symmetry breaking is then successfully triggered by the $U(1)_{B-L}$ breaking. 
Under this configuration, we have obtained the prediction for low energy observables with satisfying the SM-like Higgs boson mass to be about 125 GeV. 
The masses of the extra Higgs bosons which are mainly consist of the component fields of the triplet
are given to be around 3 TeV where the doubly-charged Higgs bosons are the heaviest among them. 
We have found that the most severe constraint is obtained for the value of $|y_\Delta^{12}y_\Delta^{11*}|\approx2.2\times10^{-4}$ from the $\mu\to 3e$ process as compared to the other combinations.

\section*{Acknowledgments}
\vspace{0.5cm}
 H.O. sincerely thank Prof. P. Ko for the great hospitality at KIAS with generous support and Prof. E.J. Chun for the nice trigger to finalize the project sophisticatedly.
 %%%
Then H.O. also expresses his sincere gratitude toward all the KIAS members, Korean cordial persons, foods, culture, weather, and all the other things. 
This work was supported by the Korea Neutrino Research Center which is established 
by the National Research Foundation of Korea(NRF) grant funded by the Korea government(MSIP) (No. 2009-0083526) (Y.O.), 
and JSPS postdoctoral fellowships for research abroad (K.Y.).

%%%%%%%%%%%%%%%%%%%%%%%%%%%%%%%%%%%

\begin{appendix}
\section{RGE}
\label{sec:rge}

In this section, we present the analytic formulae for the beta functions of all the dimensionless coupling constants at one-loop level.
The beta functions for the gauge couplings are calculated by 
\begin{align}
&(4\pi)^2\frac{dg_1}{dt} = \frac{47}{6}g_1^3, \\
&(4\pi)^2\frac{dg_2}{dt}= - \frac{5}{2}g_2^3, \\
&(4\pi)^2\frac{dg_3}{dt}= - 7 g_3^3, \\ 
&(4\pi)^2\frac{dg_{B-L}}{dt}=g_{B-L}\left(16g_{B-L}^2+\frac{44}{3}g_{B-L}g_{\text{mix}}
 +\frac{47}{6}g_{\text{mix}}^2\right), \\
&(4\pi)^2\frac{dg_{\text{mix}}}{dt}=16g_{B-L}^2g_{\text{mix}}
 +\frac{44}{3}g_{B-L}\left(g_{\text{mix}}^2+g_1^2\right)+\frac{47}{6}g_{\text{mix}}\left(g_{\text{mix}}^2+2g_1^2\right), 
\end{align}
where $g_3$ is the $SU(3)_c$ gauge coupling constant. The other coupling constants are defined in Eqs.~(\ref{kin}), (\ref{kin2}), 
(\ref{kin3}) and (\ref{kin4}). 

Those for the Yukawa couplings~\cite{Schmidt:2007nq, Okada:2014nea} are given by
\begin{align}
(4\pi)^2\frac{d y_\Delta}{dt}&=
\left(\frac12 y_D^\dag y_D+\frac12 y_\ell^\dag y_\ell^{}
+3 y_\Delta^\dag y_\Delta^{} \right)^T y_\Delta^{}
+y_\Delta^{} \left(\frac12 y_D^\dag y_D^{} +\frac12 y_\ell^\dag y_\ell^{}
+ 3 y_\Delta^\dag y_\Delta^{} \right)\nn\\
&  +\left[-\frac32\left(g_1^2+3g_2^2\right)+2\text{Tr}(y_\Delta^\dag y_\Delta^{})\right]y_\Delta^{}, \\
%%%
(4\pi)^2\frac{d y_\ell^{}}{dt}&=
y_\ell^{} \left[\frac32 y_\ell^\dag y_\ell^{} - \frac32 y_D^\dag y_D^{}+3 y_\Delta^\dag y_\Delta^{} 
+ 3 y_t^2 +\text{Tr}(y_\ell^\dag y_\ell^{} +y_D^\dag y_D^{})-\frac{15}{4} g_1^2-\frac94 g_2^2  \right] ,\\
%%%
(4\pi)^2\frac{d y_D^{}}{dt}&=
y_D^{} \left[\frac32 y_D^\dag y_D^{} +3 y_\Delta^\dag y_\Delta^{}  -\frac32 y_\ell^\dag y_\ell^{} + 3 y_t^2 
+\text{Tr}(y_\ell^\dag y_\ell^{} +y_D^\dag y_D^{}) -\frac34 g_1^2-\frac94 g_2^2  \right] ,\\
%%%
(4\pi)^2\frac{d y_t}{dt}&=
y_t \left(\frac92 y_t^2 - 8 g_3^2-\frac94 g_2^2 -\frac{17}{12}g_1^2-\frac{17}{12}g_{\text{mix}}^2
-\frac23 g_{B-L}^2 -\frac53 g_{\text{mix}}g_{B-L}^{}   \right) , \\
%%%
(4\pi)^2\frac{d y_N^{}}{dt}&= y_N^{}\left[4 y_N^\dag y_N^{}+2 \text{Tr}(y_N^\dag y_N^{})-6g_{B-L}^2\right] ,
\end{align}
where $y_t$ is the top Yukawa coupling.

Those for the scalar quartic couplings are given by
\begin{align}
(4\pi)^2\frac{d\lambda_1}{dt}&=
24\lambda_1^2 +2 \lambda_0^2  + 3 \lambda^2_4 + \lambda^2_7 
+3\lambda_4\lambda_5+\frac{5}{4}\lambda^2_5
+\frac{3}{8}\left[2g_2^4+\left(g_2^2+g_1^2+g^2_{\text{mix}}
\right)^2\right] \nn\\
&-3\lambda_1\left(  3g_2^2 + g_1^2+ g_{\text{mix}}^2  \right)
-6y^4_t + 12 \lambda_1 y^2_t, \\
%%%
(4\pi)^2\frac{d\lambda_2}{dt}&=
28\lambda_2^2 +24 \lambda_2 \lambda_3 + 6\lambda^2_3
+2 \lambda^2_4 +2\lambda_4\lambda_5 +\lambda^2_8
 +15g_2^4+12g_1^4-12g_1^2g_2^2\nn\\
 &+66 \left(2 g_{B-L} +g_{\text{mix}} \right)^4 -12\left(g_1^2+g_2^2\right)\left(2 g_{B-L} +g_{\text{mix}} \right)^2\nn\\
  &
 +4\lambda_2\left[ -6g_2^2 -3g_1^2 -3g_{\text{mix}}^2-3\left(2 g_{B-L}^{} +g_{\text{mix}} \right)^2
 +2\text{Tr}(y_\Delta^\dagger y_\Delta^{}) \right], \\
%%%
(4\pi)^2\frac{d\lambda_3}{dt}&=
18\lambda^2_3 +24\lambda_2\lambda_3 +\lambda^2_5
-6g_2^4-12g_1^4+24g_1^2 g_2^2
-4\text{Tr}(y_\Delta^{} y_\Delta^\dagger y_\Delta^{} y_\Delta^\dagger )\nn\\
 &-60 \left(2 g_{B-L} +g_{\text{mix}} \right)^4 +24\left(g_1^2+g_2^2\right)\left(2 g_{B-L} +g_{\text{mix}} \right)^2\nn\\
  &
 +4\lambda_3\left[ -6g_2^2 -3g_1^2 -3g_{\text{mix}}^2-3\left(2 g_{B-L} +g_{\text{mix}} \right)^2
 +2\text{Tr}(y_\Delta^\dagger y_\Delta^{} ) \right], \\
%%%
(4\pi)^2\frac{d\lambda_4}{dt}&=
4\lambda^2_4 +12\lambda_1\lambda_4 +16\lambda_2\lambda_4
+12\lambda_3\lambda_4 +4\lambda_1\lambda_5 +6\lambda_2\lambda_5 
+2\lambda_3\lambda_5+\lambda_5^2 +4\lambda_7\lambda_8 \nn\\
&+6g_2^4 +3g_1^4-6g_1^2g_2^2 +3g_{\text{mix}}^2\left(2 g_{B-L} +g_{\text{mix}} \right)^2
-2g_2^2g_{\text{mix}}\left(2 g_{B-L} +g_{\text{mix}} \right)
\nn\\
  &-\left[2\text{Tr}\left(y_D^T y_\Delta y_\Delta^\dagger y_D^\ast \right)
 +4\text{Tr}\left(y_\ell^T y_\Delta y_\Delta^\dagger y_\ell^\ast \right) + \text{c.c.} \right]\nn\\
  & +\lambda_4\left[ -\frac{33}{2}g_2^2 -\frac{15}{2}g_1^2-\frac{15}{2}g_{\text{mix}}^2
  -6\left(2 g_{B-L} +g_{\text{mix}} \right)^2 
  +6y_t^2+4\text{Tr}(y_\Delta^\dagger y_\Delta^{}) \right], \\
%%%
(4\pi)^2\frac{d\lambda_5}{dt}&=
\lambda_5\Big[ 4\lambda_1 +4\lambda_2 +8\lambda_3 +8\lambda_4+4\lambda_5 \nn\\ & 
 -\frac{33}{2}g_2^2 -\frac{15}{2}(g_1^2+g_{\text{mix}}^2)-6\left(2 g_{B-L} +g_{\text{mix}} \right)^2+6y_t^2 +4\text{Tr}(y_\Delta^\dagger y_\Delta^{})  \Big] \nn\\
&+8\lambda_0^2 +12g_1^2g_2^2 +12g_2^2g_{\text{mix}}\left(2 g_{B-L} +g_{\text{mix}} \right)
-\left[ 4\text{Tr}\left(y_\ell^T y_\Delta y_\Delta^\dagger y_\ell^\ast \right) + \text{c.c.} \right], \\
%%%
(4\pi)^2\frac{d\lambda_6}{dt}&=
20\lambda_6^2 +2 \lambda^2_7 + 3\lambda^2_8 
+96 g^4_{B-L} -48\lambda_6 g^2_{B-L}
-8\text{Tr}(y_N^\dagger y_N y_N^\dagger y_N^{}) 
+ 8 \lambda_6 \text{Tr}(y_N^\dagger y_N^{} ), \label{lam6}\\
%%%
(4\pi)^2\frac{d\lambda_7}{dt}&=
4\lambda^2_7+12\lambda_1\lambda_7 +8\lambda_6\lambda_7 +6\lambda_4\lambda_8
+3\lambda_5\lambda_8+12\lambda_0^2 \nn\\
&+\lambda_7\left[ -\frac{9}{2}g_2^2 -\frac{3}{2}g_1^2-\frac32 g_{\text{mix}}^2-6\left(2 g_{B-L} +g_{\text{mix}} \right)^2
 +6y_t^2+8\text{Tr}(y_N^\dagger y_N^{})\right], \\
%%%
(4\pi)^2\frac{d\lambda_8}{dt}&=
4\lambda_8^2 +\lambda_7\left( 4\lambda_4 +2\lambda_5 \right)
+\lambda_8 \left( 16\lambda_2 +12\lambda_3 +8\lambda_6\right)
+4\lambda_0^2 +96 g_{B-L}^4\nn\\
&
+\lambda_8\left[-12g_2^2 -6g_1^2 -6g_{\text{mix}}^2-6\left(2 g_{B-L} +g_{\text{mix}} \right)^2
 +4\text{Tr}(y_\Delta^\dagger y_\Delta^{} )
+8\text{Tr}(y_N^\dagger y_N^{}) \right], \\
%%%
(4\pi)^2\frac{d\lambda_0}{dt}&=
\lambda_0 \Big[ 4\lambda_1 +4\lambda_4 +6\lambda_5 +4\lambda_7 +2\lambda_8
 -\frac{21}{2}g_2^2 -\frac{9}{2}g_1^2-3\left(2 g_{B-L} +g_{\text{mix}} \right)^2
\nn\\
  & +4\text{Tr}(y_N^\dagger y_N^{} )  +2\text{Tr}(y_\Delta^\dagger y_\Delta^{} ) +6y_t^2 \Big]
  -4\left[ \text{Tr}(y_N y_D^\dagger y_\Delta^\ast y_D^\ast ) +\text{c.c.} \right]. 
\end{align}
%%%

\end{appendix}

\end{document}